\documentclass[aps,pra,reprint,amsmath,amsfonts,amssymb,superscriptaddress]{revtex4-1}
\usepackage{graphicx,float,calc}
\usepackage{color,bm}
\usepackage{ulem}
\usepackage{braket}
\usepackage[colorlinks,urlcolor=blue,citecolor=blue,linkcolor=blue]{hyperref}

\begin{document}
	
\title{Topological Anderson insulators in two-dimensional non-Hermitian disordered systems}

\author{Ling-Zhi Tang}
\author{Ling-Feng Zhang}
\affiliation{Guangdong Provincial Key Laboratory of Quantum Engineering and Quantum Materials, GPETR Center for Quantum Precision Measurement and SPTE, South China Normal University, Guangzhou 510006, China}
\author{Guo-Qing Zhang}\thanks{zhangptnoone@m.scnu.edu.cn}
\author{Dan-Wei Zhang}\thanks{danweizhang@m.scnu.edu.cn}
\affiliation{Guangdong Provincial Key Laboratory of Quantum Engineering and Quantum Materials, GPETR Center for Quantum Precision Measurement and SPTE, South China Normal University, Guangzhou 510006, China}
\affiliation{Frontier Research Institute for Physics, South China Normal University, Guangzhou 510006, China}

\begin{abstract}
The interplay among topology, disorder, and non-Hermiticity can induce some exotic topological and localization phenomena. Here we investigate this interplay in a two-dimensional non-Hermitian disordered {\color{black} Chern-insulator model with two typical kinds of non-Hermiticities}, the nonreciprocal hopping and on-site gain-and-loss effects. The topological phase diagrams are obtained by numerically calculating two topological invariants in the real space, which are the disorder-averaged open-bulk Chern number and the generalized Bott index, respectively. We reveal that the nonreciprocal hopping (the gain-and-loss effect) can enlarge (reduce) the topological regions and the topological Anderson insulators induced by disorders can exist under both kinds of non-Hermiticities. Furthermore, we study the localization properties of the system in the topologically nontrivial and trivial regions by using the inverse participation ratio and the expansion of single particle density distribution.
\end{abstract}

\date{\today}

\maketitle

\section{Introduction}
Topological insulators, as a new class of states of matter with nontrivial band structures, have witnessed fast development in condensed matter physics \cite{Hasan2010,XLQi2011} and tunable engineered systems \cite{DWZhang2018,Cooper2019,Goldman2016,Schroer2014,Roushan2014,XTan2018,XTan2019b,Lee2018b,Huber2016,LLu2014,Ozawa2019} in recent years. Topological insulators are generally characterized by topological invariants of extended bulk states and gapless edge states under open boundary conditions (OBCs). For instance, two-dimensional (2D) Chern insulators are topologically characterized by the Chern number and chiral edge states \cite{XLQi2011}. These topological characters are robust against certain types of weak disorders as the topological band gap is preserved under these perturbations. For sufficiently strong disorders, the systems usually become trivial insulators as the band gaps close and all states are localized due to the Anderson localization \cite{Anderson1958}. In contrast to this common wisdom, a counter-intuitive behavior that the disorder can induce the non-trivial phase from trivial phase has been revealed and the disorder-induced topological phase is known as the topological Anderson insulator (TAI) \cite{JLi2009}. TAIs have been theoretically investigated in various models and systems \cite{JLi2009,Groth2009,HJiang2009,HMGuo2010,Altland2014,Mondragon-Shem2014,Titum2015,BLWu2016,Sriluckshmy2018,ZQZhang2019,JHZheng2019,Kuno2019,ZQZhang2019,RChen2019}, such as the disordered Su-Schrieffer-Heeger (SSH) chain \cite{WPSu1979} and Haldane model \cite{Haldane1988} of Chern insulators.
Recently, the experimental observation of TAIs in two engineered systems with tunable disorder and topology was reported, i.e., one-dimensional (1D) cold atomic wires \cite{Meier2018} and 2D photonic waveguide arrays \cite{Stutzer2018}.

Most studies of topological states focus on Hermitian systems; however, the recent theoretical and experimental advances of non-Hermitian physics have inspired the extension of topological systems to the non-Hermitian regime \cite{Bender1998,El-Ganainy2018,Miri2019}. A remarkable, fast-growing effort has been undertaken to explore novel topological states and phenomena in non-Hermitian systems \cite{Rudner2009,Zeuner2015,Esaki2011,YHu2011,BZhu2014,Malzard2015,Leykam2017,YXu2017b,XZhan2017,Lee2016,Fring2016,QBZeng2017,Jin2017,SYao2018,FSong2019,Kunst2018,YXiong2018,Yuce2018,HJiang2018,LWZhou2018,RWang2018,LJin2019,Borgnia2019,ZGong2018,SYao2018b,HShen2018,Takata2018,YChen2018,LJLang2018,Harari2018,Bandres2018,HZhou2018,TSDeng2019,Ezawa2019,Kawabata2019,Ghatak2019,Kawabata2018,Yuce2019,TLiu2019,Lee2019,Luitz2019,Yamamoto2019,Yuce2018,CHYin2018, Herviou2019,DWZhang2020,TLiu2020,ZXu2020}, where the non-Hermiticites include the non-Hermitian gain and loss \cite{LFeng2017,El-Ganainy2018,Miri2019}, the nonreciprocal hopping \cite{Hatano1996,Hatano1997}, and the dissipation in open systems \cite{Diehl2008}. It has also been revealed that non-Hermitian disordered systems have unique localization properties \cite{Hatano1996,Hatano1997,Carmele2015,Levi2016,Hamazaki2019,Mejia-Cortes2015,QBZeng2017,Alvarez2018,PWang2019}. Moreover, the interplay of topology and disorder in 1D non-Hermitian systems has been studied \cite{DWZhang2019,XWLuo2019,Longhi2019,HJiang2019,QBZeng2020,JHou2019}. The topological phase (transition) in 1D non-Hermitian
quasicrystals has been explored \cite{Longhi2019,HJiang2019,QBZeng2020,JHou2019}. Notably, non-Hermitian TAIs induced by a combination of disorder and non-Hermiticity in the SSH chains consisting of nonreciprocal disordered hoppings have been uncovered \cite{DWZhang2019,XWLuo2019}.

{\color{black} In this work, we explore the topological and localization properties in a 2D non-Hermitian disordered Chern-insulator model, which in the Hermitian limit is a square-lattice variation of the Haldane model \cite{Haldane1988}.} Two typical kinds of non-Hermiticities from the nonreciprocal hopping and the Zeeman potential with gain-and-loss effects are considered in the constructed 2D tight-binding model. We obtain the topological phase diagrams by numerically calculating two topological invariants in real space, which are the disorder-averaged open-bulk Chern number and generalized Bott index \cite{Loring2010,HHuang2018,QBZeng2020}. In the Hermitian limit, we show that the disorder in the system can induce a Chern-insulator phase. We demonstrate that the nonreciprocal hopping (the gain-and-loss effect) can enlarge (reduce) the topological regions and find that topological Anderson insulators induced by disorders can exist under these two kinds of non-Hermiticities. Moreover, we study the localization properties of the system in the topologically nontrivial and trivial regions by using the inverse participation ratio and the expansion of the single particle density distribution.

The rest of this paper is organized as follows. We first propose the 2D non-Hermitian disordered Chern-insulator model with nonreciprocal hopping and on-site gain-and-loss terms, and then introduce two approaches for calculating the topological invariants for the non-Hermitian disordered systems in Sec.~\ref{sec2}. Section~\ref{sec3} is devoted to investigating topological phase diagrams and discussing the non-Hermtian TAI in the 2D system. We further study the localization properties of our model in Sec.~\ref{sec4}. A conclusion and discussion are finally presented in Sec.~\ref{sec5}.

\section{\label{sec2}model and topological invariants}

\begin{figure}
	\centering
	\includegraphics[width=0.38\textwidth]{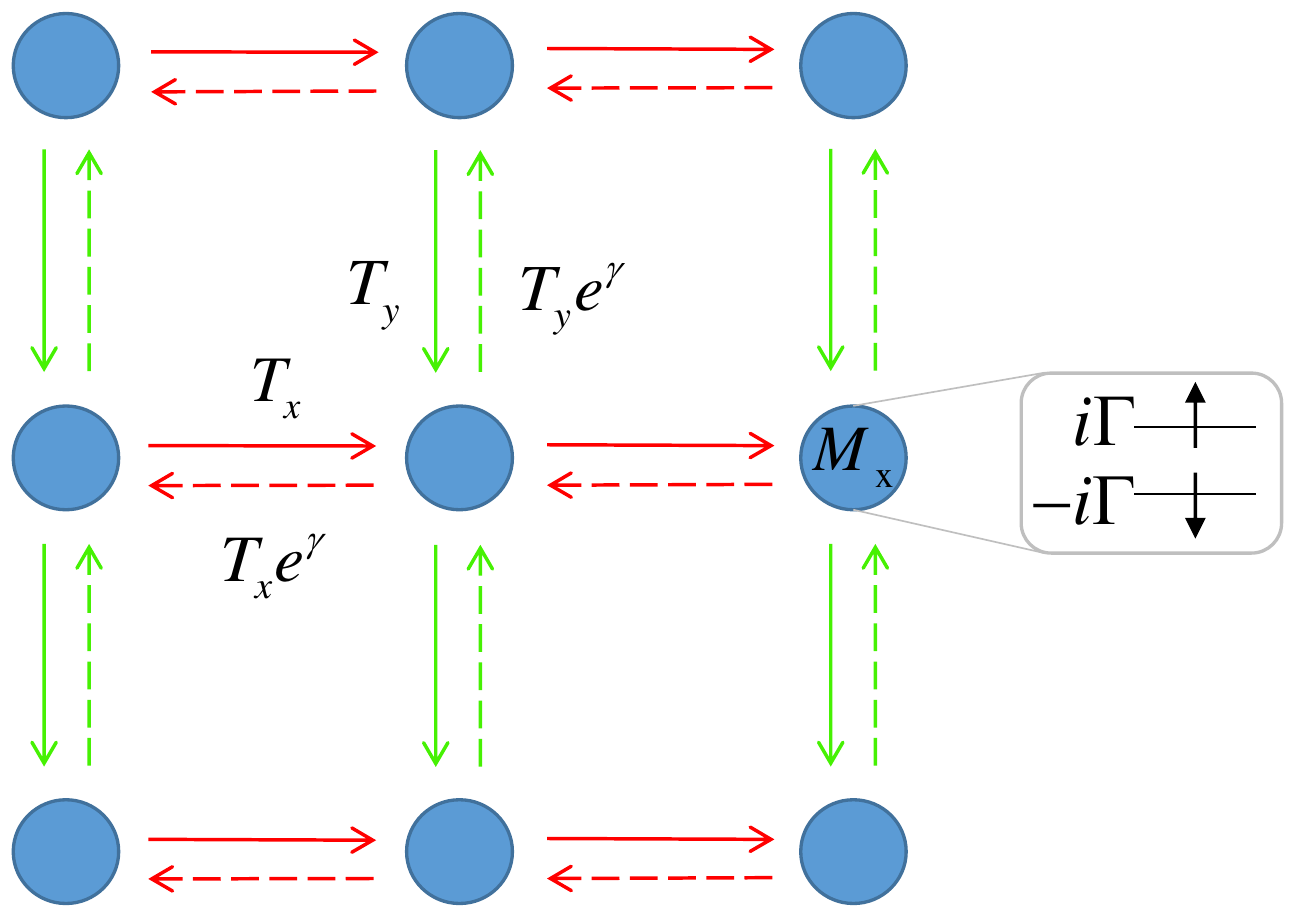}
	\caption{(Color online) Schematic 2D lattice of the non-Hermitian disordered Chern-insulator model with the site index $\mathbf{x}=(x,y)$. Here $T_{x,y}$ and $T_{x,y}e^\gamma$ denote the nonreciprocal intercell hopping matrices on the (pseudo-)spin basis $\{\uparrow,\downarrow\}$ along the $x,y$ axis, and $M_\textbf{x}$ denotes the on-site potential with a disordered Zeeman term $m_\textbf{x}\sigma_z$ and the gain-and-loss term $i\Gamma\sigma_z$ acting on the (pseudo-)spins.
	}\label{fig1}
\end{figure}

We begin by constructing a {\color{black} 2D non-Hermitian square-lattice Chern-insulator model}, where the non-Hermiticity arises from the nonreciprocal hopping and the Zeeman potential with on-site gain-and-loss term. As depicted in Fig. \ref{fig1}, the tight-binding model reads
\begin{equation}\label{H}
\begin{aligned}
H=\sum_\textbf{x}\sum_{j=x,y}[c_\textbf{x}^\dagger T_j c_{\textbf{x}+\textbf{e}_j}+c_{\textbf{x}+\textbf{e}_j}^\dagger T_j^\dagger e^\gamma c_\textbf{x}]+\sum_\textbf{x}c_\textbf{x}^\dagger M_\textbf{x} c_\textbf{x},
  \end{aligned}
\end{equation}
where $c_\textbf{x}^\dagger=(c_{\textbf{x},\uparrow}^\dagger ,c_{\textbf{x},\downarrow}^\dagger)$ is a two-component creation operator that creates a particle at lattice site $\textbf{x}=(x,y)$ with (pseudo-)spin $\{\uparrow,\downarrow\}$; $c_\textbf{x}$ is the corresponding annihilation operator; $T_j=-\frac{1}{2}t_j\sigma_z-\frac{i}{2}\nu_j\sigma_j$; $M_\textbf{x}=(m_\textbf{x}+i\Gamma)\sigma_z$, with $\gamma$ and $\Gamma$ being the non-Hermiticity strengths of the nonreciprocal intercell hopping and the gain-and-loss term; $\textbf{e}_j$ is the unit vector along the $j$ direction ($j=x,y$), and $\sigma_j$ are Pauli matrices on the spin basis. The hopping strengths $t_x=t_y=t=1$ and $v_x=v_y=t=1$ are set with $t=1$ as the energy unit hereafter for convenience. In this model, we consider the diagonal disorder in the Zeeman potential $m_\textbf{x}=m+W\omega_\textbf{x}$, where $m$ is a constant Zeeman strength and $\omega_\textbf{x}$ denotes independent random numbers chosen uniformly in the range $[-1,1]$ with the disorder strength $W$.

{\color{black} In the Hermitian and clean limit of $\gamma=\Gamma=W=0$, the Hamiltonian in Eq. (\ref{H}) returns to the square-lattice model of 2D Chern (quantum Hall) insulators \cite{XLQi2011}, which is a variation of the Haldane model \cite{Haldane1988}. In this clean case, when $|m|<2$, the band topology can be characterized by a non-zero Chern number, and the system has chiral edge modes located at the boundary of the square lattice under OBCs. When $|m|>2$, the two energy bands become trivial with a zero Chern number and the edge modes disappears. Thus, a topological phase transition between Chern insulators and trivial insulators occurs at $|m|=t_x+t_y=2$.}

In the presence of disorder, the translation symmetry is broken and the wave vector is no longer a good quantum number. In this case, the conventional Chern number whose formula is defined in terms of Bloch vectors is inapplicable. In addition, due to the non-Hermitian skin effect when $\gamma\neq0$ \cite{SYao2018,FSong2019,Kunst2018}, the conventional bulk-boundary correspondence based on the Hermitian model may fail to correctly describe the relation between the number of chiral edge states and the conventional Chern number. Here, we use the open-bulk Chern number developed in Ref. \cite{FSong2019} to characterize the topological properties of our non-Hermitian disordered model, which is calculated in real space and can recover to the Hermitian (clean) case \cite{Bianco2011}. For a disorder configuration denoted by $s$, the corresponding open-bulk Chern number is given by \cite{FSong2019}
\begin{equation}
C_s=\frac{2\pi i}{L_x^\prime L_y^\prime}\text{Tr}^\prime(P_s[[X,P_s],[Y,P_s]]),
\end{equation}
where $X$ and $Y$ denote the coordinate operators, $P_s$ is a valence-band projector operator, $\text{Tr}^\prime$ stands for the trace within the central region of size $L_x^\prime \times L_y^\prime$ with system size $L_x\times L_y$ and $L_j^\prime=L_j-2l_j$. Here $l_j$ should be sufficiently large to dissolve the finite-size effect. For convenience, we set $L_x=L_y=L$ in our numerical simulations. In the biorthonormal basis, $P_s$ is calculated under OBCs,
\begin{equation}
P_s=\sum|nR\rangle_s {_s}\langle nL|,
\end{equation}
where $\ket{nR}_s$ and $\ket{nL}_s$ represent the $n$-th right and left eigenstates obtained from the eigenfunctions $H_s\ket{nR}_s=E_{s,n}\ket{nR}_s$ and $H_s^\dagger\ket{nL}_s=E_{s,n}^\ast\ket{nL}_s$, respectively. Note that the eigenstates obey the biorthonormal condition ${_s}\langle nL|n^\prime R\rangle_s=\delta_{nn^\prime}$ and $C_s$ converges to an integer when $L\rightarrow\infty$. We define the disorder-averaged open-bulk Chern number as
\begin{equation}\label{C}
C=\frac{1}{N_s}\sum_{s=1}^{N_s}C_s,
\end{equation}
where $N_s$ denotes the disorder configuration number and is set as $N_s\in[100,300]$ to make $C$ converge.

We also consider the Bott index as another topological invariant \cite{Loring2010,HHuang2018,QBZeng2020}, which can be calculated in real space and generalized for non-Hermitian models. The Bott index measures the commutativity of two unitary and commuting matrices, and also discriminates pairs of commuting matrices from those far from commuting. The Bott index equals the Chern number in the thermodynamic limit. For the disorder configuration, the generalized Bott index is given by \cite{Loring2010,HHuang2018,QBZeng2020}
\begin{equation}
B_s=\frac{1}{2\pi}\text{Im}\left[\text{Tr}\left(\text{ln}V_sU_sV_s^\dagger U_s^\dagger\right)\right],
\end{equation}
where $U_s$ and $V_s$ are the projected position operators and the matrix elements
\begin{equation}
U_{s,mn}={_s}\langle mL|e^{2\pi iX/L}\ket{nR}_s
\end{equation}
\begin{equation}
V_{s,mn}={_s}\langle mL|e^{2\pi iY/L}\ket{nR}_s
\end{equation}
are defined in the biorthonormal basis for non-Hermitian systems. The disorder-averaged Bott index is given by
\begin{equation}\label{B}
B=\frac{1}{N_s}\sum_{s=1}^{N_s}B_s.
\end{equation}
When $\gamma=\Gamma=0$, the Chern number in Eq. (\ref{C}) and the Bott index in Eq. (\ref{B}) reduce to those in Hermitian systems with $\ket{nR}_s=\ket{nL}_s$ involved in the calculations.

\section{\label{sec3}Topological phase diagrams}

\begin{figure}
	\centering
	\includegraphics[width=0.45\textwidth]{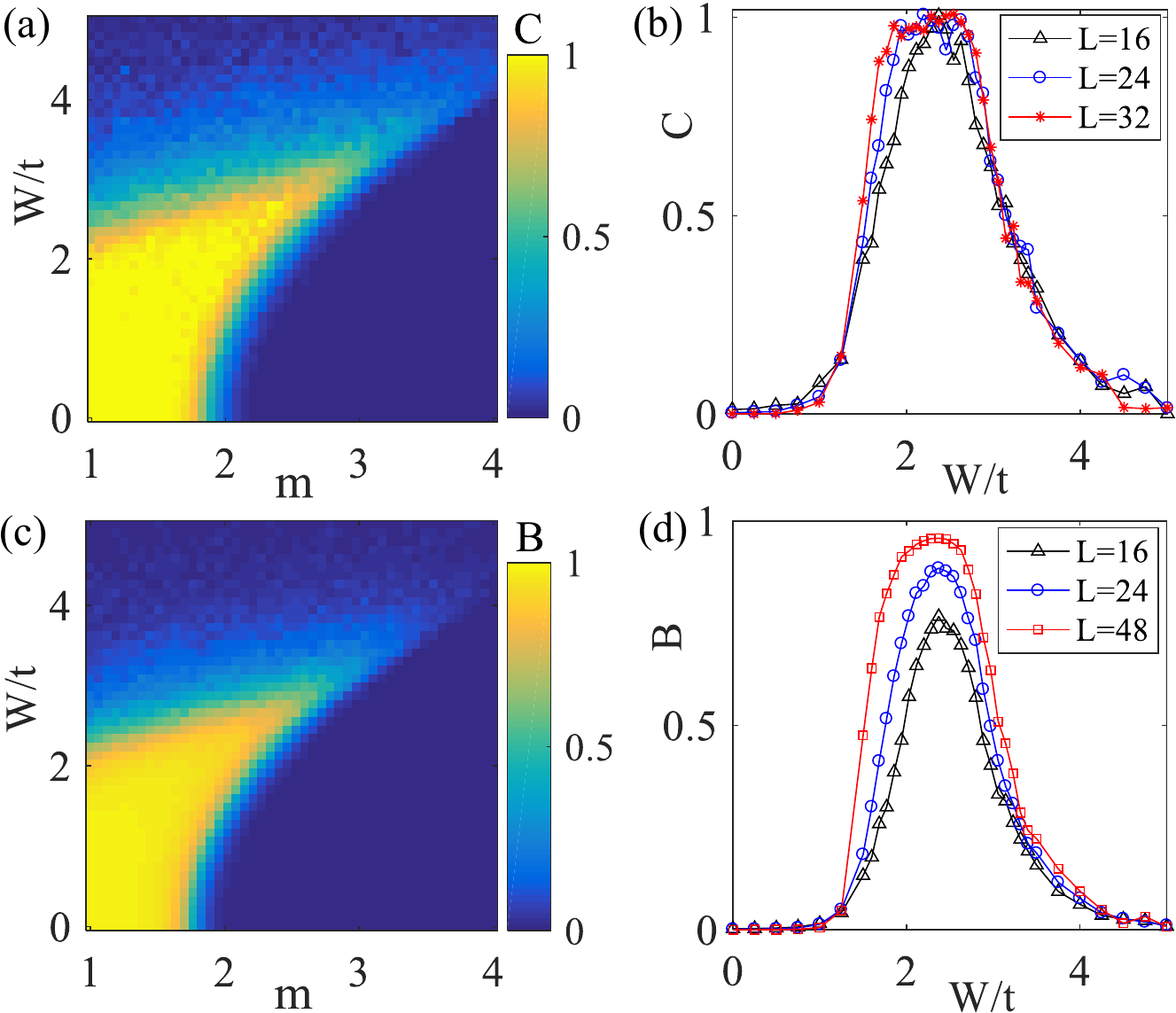}
	\caption{(Color online) {\color{black}Hermitian case.} (a) The Chern number $C$ and (c) Bott index $B$ as a function of the potential $m$ and the disorder strength $W$ for a system size $L=16$. (b) The Chern number $C$ and (d) Bott index $B$ as a function of $W$ for $m=2.2$ and various system sizes $L$. $\gamma=\Gamma=0$ for all panels.
	}\label{fig2}
\end{figure}

We first consider the disorder effect on the topological phase of the 2D Hermitian Chern-insulator model with $\gamma=\Gamma=0$ in Eq. (\ref{H}). The obtained topological phase diagram in Fig. \ref{fig2}(a) shows the disorder-averaged Chern number $C$ as a function of the potential strength $m$ and the disorder strength $W$ for $L=16$. In the clean limit $W=0$, a topological transition between the non-trivial phase with $C=1$ and the trivial phase with $C=0$ occurs at $m=2$ as excepted. As the disorder strength grows to a modest region, the topological transition point moves forward larger $m$, and the region of Chern insulators is enlarged. Remarkably, in the region $m\in[2,3]$, the topological phase can be induced by moderate disorder from the trivial phase in the clean case $W=0$, which indicates the TAI in this 2D Hermitian system. A consistent topological phase diagram with the TAI is obtained from the calculated Bott index $B$, as shown in Fig. \ref{fig2}(c). Notably, $C$ and $B$ can approach to 1 in the TAI region by increasing the lattice size, with two examples as a function of $W$ with $m=2.2$ shown in Figs. \ref{fig2}(b) and \ref{fig2}(d).

\begin{figure}
	\centering
	\includegraphics[width=0.48\textwidth]{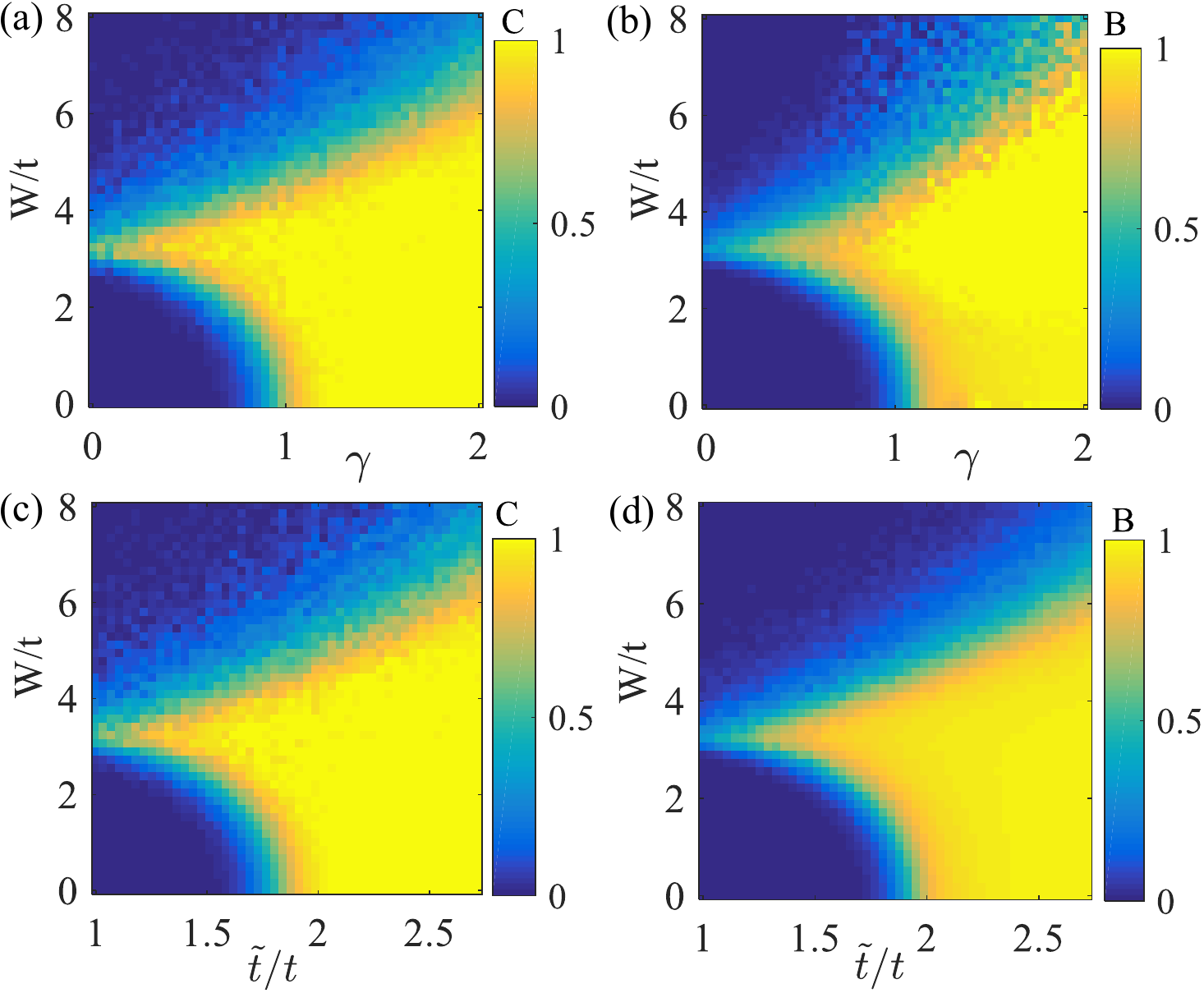}
	\caption{(Color online) {\color{black} Non-Hermitian case with nonreciprocal hopping.} (a) The Chern number $C$ and (b) Bott index $B$ as a function of the nonreciprocal parameter $\gamma$ and the disorder strength $W$ for $\Gamma=0$, $m=3$ and $L=16$.  (c) The Chern number $C$ and (d) Bott index $B$ as a function of $W$ and $\tilde{t}$ after the similarity transformation of the non-Hermitian Hamiltonian $H$ with $\Gamma=0$ under OBCs.
	}\label{fig3}
\end{figure}

We now study the effect of the combined disorder and non-Hermiticity on an initially trivial phase in the clean and Hermitian cases. We consider the first kind of non-Hermiticity from the nonreciprocal hopping $\gamma$ with fixed $\Gamma=0$ and $m=3$. In Figs. \ref{fig3}(a) and \ref{fig3}(b), we present the numerically calculated Chern number $C$ and Bott index $B$ as a function of $\gamma$ and disorder strength $W$, respectively. As shown in the topological phase diagrams, the non-Hermiticity $\gamma$ can induce a topological phase transition from the trivial phase to the topological phase in the clean case of $W=0$ and at modest disorder $W\lesssim6$. Thus, the nonreciprocal hopping can enlarge the topological region in this 2D system. More interestingly, the 2D non-Hermitian TAI can be induced by the combination of disorder and non-Hermiticity \cite{DWZhang2019,XWLuo2019}, as an extension of the TAI in Hermitian systems \cite{JLi2009,Groth2009,HJiang2009,HMGuo2010,Altland2014,Mondragon-Shem2014,Titum2015,BLWu2016,Sriluckshmy2018,ZQZhang2019,JHZheng2019,Kuno2019,ZQZhang2019,RChen2019}.

To further understand the topological phase diagrams in Figs. \ref{fig3}(a) and \ref{fig3}(b), we take a similarity transformation of the non-Hermitian Hamiltonian in Eq. (\ref{H}) with $\Gamma=0$ for a given configuration $H_s$ under OBCs $\tilde{H}_s=S^{-1}H_sS$, where the transformation matrix takes the diagonal form $S=\text{diag}(1,1,r,r,r,r,\cdots,r^{L^2/2-1},r^{L^2/2-1},r^{L^2/2-1},r^{L^2/2-1},\\r^{L^2/2},r^{L^2/2})$ with $r=\sqrt{e^\gamma}$. After the similarity transformation, one has the eigenfunction for a Hermitian model $\tilde{H}_s\ket{\tilde{n}}_s=E_{s,n}\ket{\tilde{n}}_s$, with the wave function $\ket{\tilde{n}}_s=S^{-1}\ket{nR}_s$ and the corresponding Hermitian Hamiltonian $\tilde{H}_s$, with parameters $\tilde{t}_j=\sqrt{e^\gamma}t_j$, $\tilde{\nu}_j=\sqrt{e^\gamma}\nu_j$, and $\tilde{m}_\textbf{x}=m_\textbf{x}$. In this case, $H_s$ and $\tilde{H}_s$ have the same real energy spectrum $E_{s,n}$ under OBCs, which is confirmed in our numerical simulations. By using the Hermitian Hamiltonian $\tilde{H}_s$, we calculate the corresponding disorder-averaged Chern number and Bott index as a function of $\tilde{t}$ ($\tilde{t}_x=\tilde{t}_y=\tilde{t}$) and $W$, as shown in Figs. \ref{fig3}(c) and \ref{fig3}(d). One can find that the topological phase boundaries of the Hermitian model is consistent with the original non-Hermitian model under the similarity transformation. In this non-Hermitian case, the TAIs can be topologically connected to those in the Hermitian case through the similarity transformation under OBCs. However, as is clearly seen from the transformation, the wave functions (the density distribution) of the bulk states are accumulated at the boundaries, which is the non-Hermitian skin effect ~\cite{SYao2018,FSong2019,Kunst2018}.

\begin{figure}
	\centering
	\includegraphics[width=0.48\textwidth]{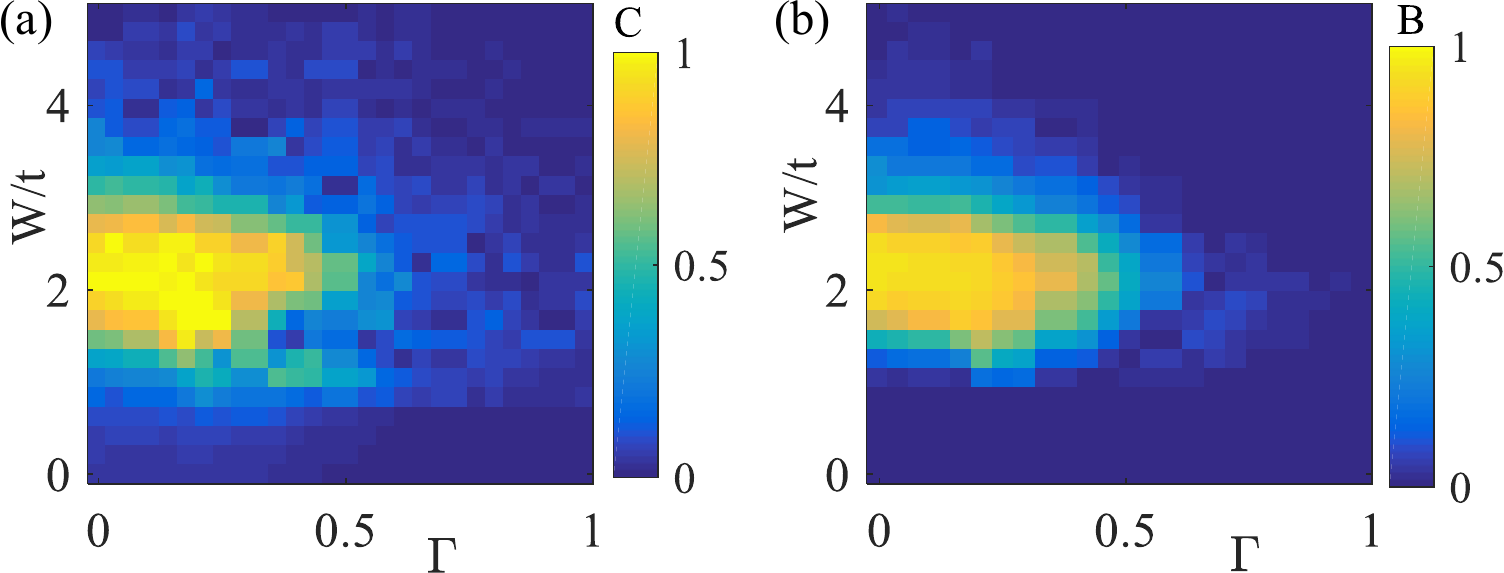}
	\caption{(Color online) {\color{black}Non-Hermitian case with gain and loss.} (a) The Chern number $C$ and (b) Bott index $B$ as a function of the non-Hermitian gain-and-loss parameter $\Gamma$ and the disorder strength $W$ for $\gamma=0$, $m=2.2$ and $L=16$.
	}\label{fig4}
\end{figure}

We also consider the non-Hermitian gain and loss in the model with $\Gamma\neq0$ and $\gamma=0$. In this case, the energy spectrum is generally complex, and there is no similarity transformation to Hermitian models. We find that the increasing of the gain-and-loss strength can not enlarge and generally reduces the topological regions in the topological phase diagrams. However, the disorder-induced TAI can persist for small $\Gamma$. Figures. \ref{fig4}(a) and \ref{fig4}(b) show the calculated $C$ and $B$ as a function of $\Gamma$ and $W$ for $m=2.2$ and $L=16$. In the Hermitian limit $\Gamma=\gamma=0$, the TAI exists for modest $W$. When $\Gamma$ increases, the TAI persist under finite non-Hermitian gain and loss and finally becomes trivial when $\Gamma\gtrsim0.5$. Since the nonreciprocal hopping enlarges the topological region, one can except the TAI to exist in the general non-Hermitian cases with non-vanishing $\gamma$ and $\Gamma$.

\section{\label{sec4}Localization properties}
\begin{figure}
	\centering
	\includegraphics[width=0.48\textwidth]{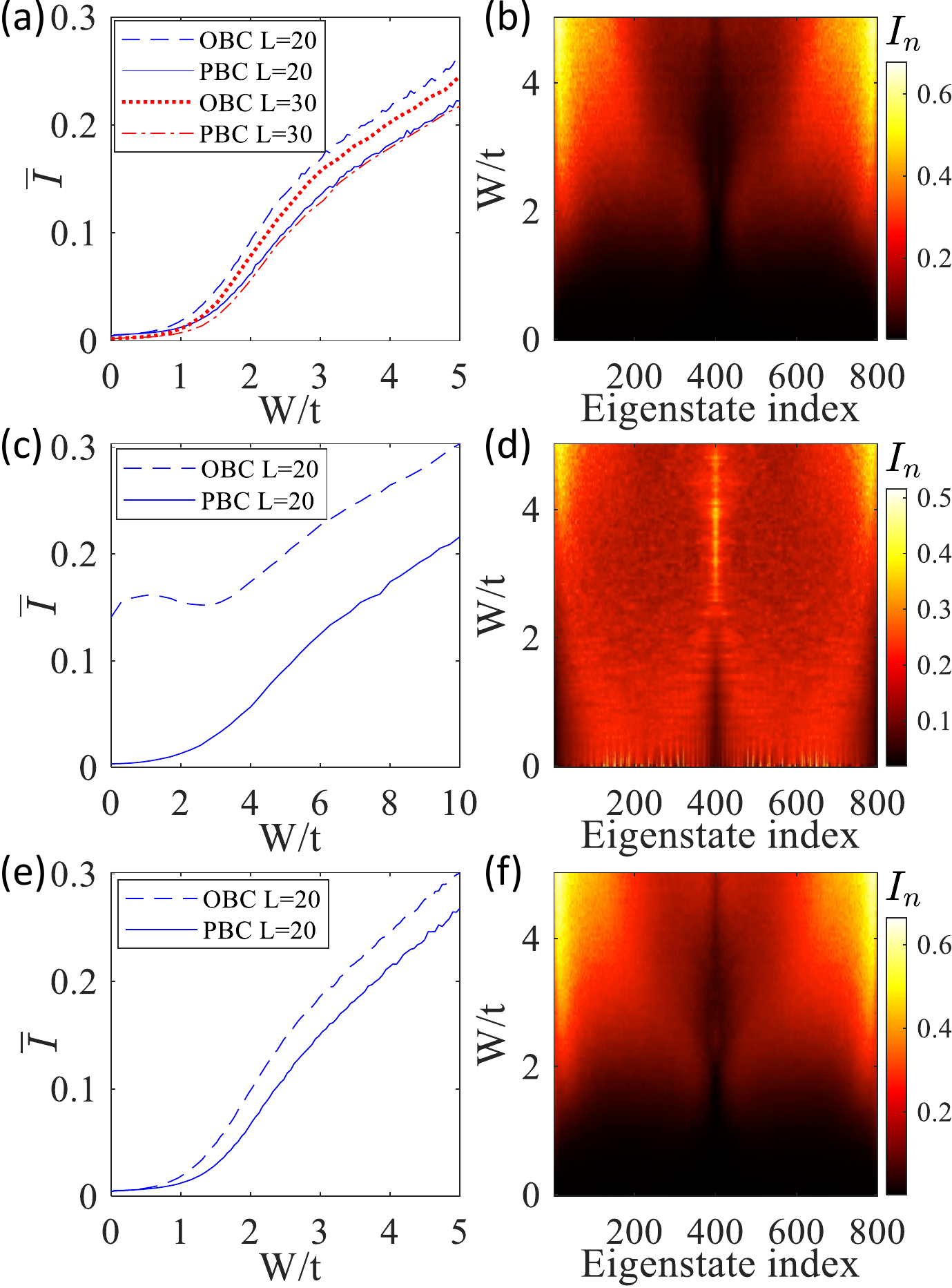}
	\caption{(Color online) {\color{black} Disorder-averaged IPR of the right eigenstates for the system size $L=20$.} (a) and (b) The Hermitian limit of the model with $\gamma=0$, $\Gamma=0$ and $m=2.2$. (c) and (d) The nonreciprocal case of the non-Hermitian model with $\gamma=0.75$, $\Gamma=0$ and $m=3$. (e) and (f) The gain-and-loss case of the non-Hermitian model with $\gamma=0$, $\Gamma=0.25$ and $m=2.2$. (a), (c) and (e) Blue dashed lines and blue solid lines denote $\bar{I}$ as a function of the disorder strength $W$ under OBCs and PBCs, respectively. The red dotted line and red dot-dashed line in (a) denote $\bar{I}$ for $L=30$. (b),(d) and (f) $I_n$ in the whole energy spectrum as a function of $W$.
	}\label{fig5}
\end{figure}

In this section, we consider the localization properties of the non-Hermitian systems induced by the disorder. {\color{black} First, we calculate the disorder-averaged inverse participation ratio (IPR)~\cite{PhysRevLett.84.3690,edelman2005random}. For non-Hermitian systems, the IPR can be directly defined for the right (left) eigenstates governed by the system Hamiltonian $H$ ($H^{\dag}$) \cite{Mejia-Cortes2015,QBZeng2017,Alvarez2018,PWang2019,DWZhang2019,Longhi2019,HJiang2019,QBZeng2020,JHou2019} or under the biorthogonal basis from both right and left eigenstates \cite{ZGong2018,PWang2019,XWLuo2019}. The disorder-averaged IPR of the $n$-th right eigenstate $\ket{nR}_s$ is defined as
\begin{equation}\label{IPR}
I_n=\overline{\frac{\sum_{\textbf{x}}|\phi_\textbf{x}^{(s,n)}|^2}{(\sum_{\textbf{x}}|\phi_\textbf{x}^{(s,n)}|)^2}},
\end{equation}
where $\phi_\textbf{x}^{(s,n)}={_s}\bra{nR}\textbf{\^{x}}\ket{nR}_s$ with $\textbf{\^{x}}$ being the position operator of the lattice site $\textbf{x}$, and the overline indicates the average over many disorder configurations of index $s$. Then the mean IPR $\bar{I}=\frac{1}{N}\sum_{n=1}^NI_n$, with $N=2L^2$, is given by averaging over the whole energy spectrum. The value of IPR saturates to a finite value for a localized state in this 2D system but scales as $L^{-2}$ and approaches zero in the thermodynamic limit for a delocalized state.

Because $I_n$ in Eq. (\ref{IPR}) measures the eigenstate density distribution without the non orthogonality of different right eigenstates, the IPR can be used to capture the localization properties of the right eigenstates of the non-Hermitian Hamiltonian $H$ in Eq. (\ref{H}) (see examples for 1D non-Hermitian systems in Refs. \cite{Mejia-Cortes2015,QBZeng2017,Alvarez2018,PWang2019,DWZhang2019,Longhi2019,HJiang2019,QBZeng2020,JHou2019}). Furthermore, one can define the biorthogonal IPR [see Eq. (\ref{BIPR})] from both right and left eigenstates to study the non-Hermitian effect on the localization properties of non-Hermitian disordered systems \cite{ZGong2018,PWang2019,XWLuo2019}.}

In the Hermitian limit of this model with $\gamma=0$, $\Gamma=0$ and $m=2.2$, Fig. \ref{fig5}(a) shows the corresponding $\bar{I}$ as a function of $W$ under both OBCs and periodic boundary conditions (PBCs). The value of $\bar{I}$ increases with the enhancement of the disorder strength $W$, which implies that the initial extended bulk states are gradually localized by the growing disorder in both boundary conditions and totally localized in the TAI region. Although $\bar{I}$s under PBCs are less sensitive to system size $L$ than those under OBCs, $\bar{I}$s under these two boundary conditions approach each other in the large-$L$ limit. $I_n$s as a function of $W$ in the whole energy spectrum under OBCs are plotted in Fig. \ref{fig5}(b). It is clear that the eigenstates in the middle of the spectrum show small values of the IPR and these topologically protected zero-mode eigenstates are robust to disorder.

\begin{figure}
	\centering
	\includegraphics[width=0.48\textwidth]{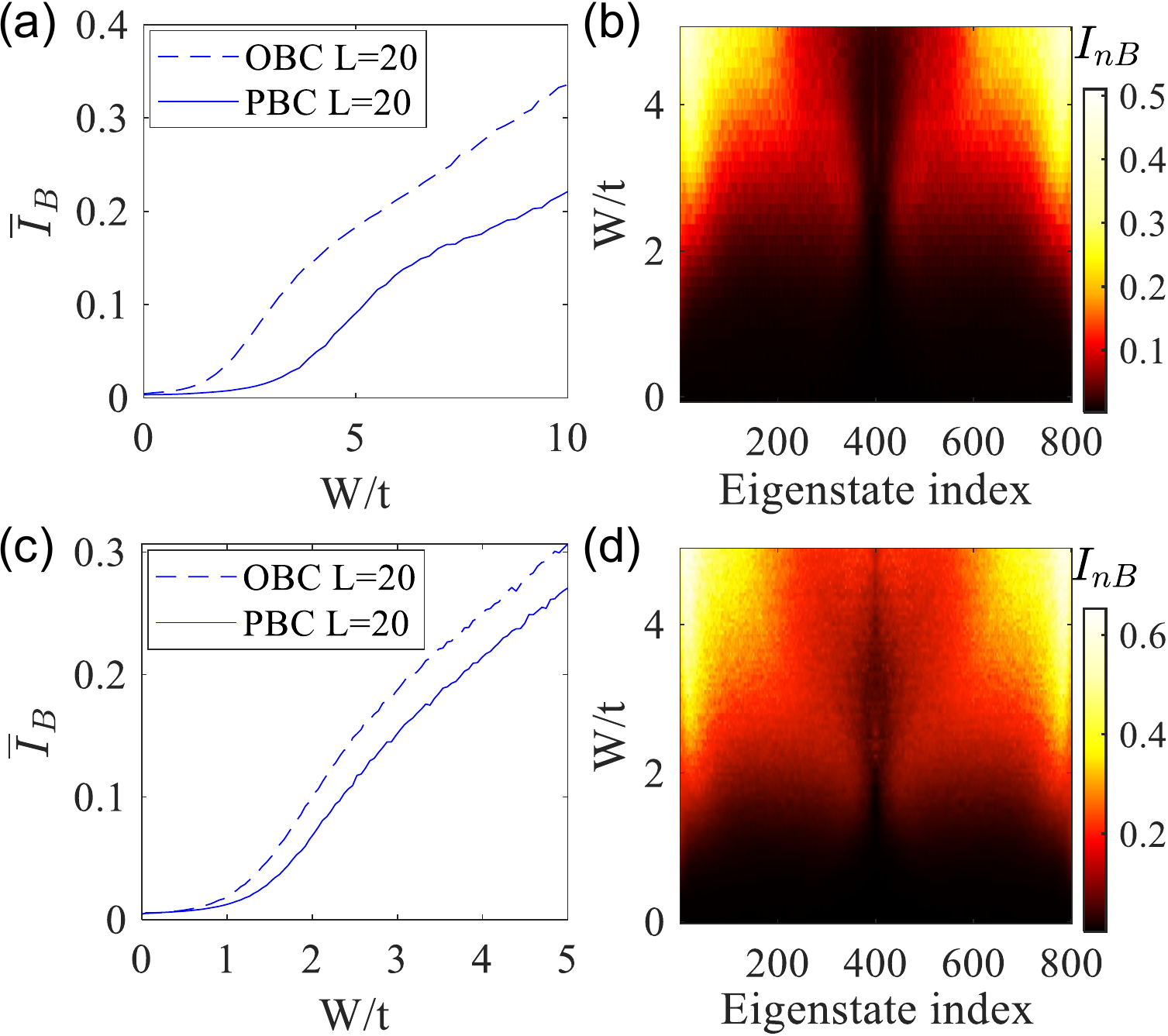}
	\caption{(Color online) {\color{black} {\color{black} Disorder-averaged biorthogonal IPR for system size $L=20$.} (a) and (b) The nonreciprocal case of the non-Hermitian model with $\gamma=0.75$, $\Gamma=0$, and $m=3$. (c) and(d) The gain-and-loss case of the non-Hermitian model with $\gamma=0$, $\Gamma=0.25$, and $m=2.2$. (a) and (c) Blue dashed lines and blue solid lines denote $\bar{I}_B$ as a function of the disorder strength $W$ under OBCs and PBCs, respectively. (b) and (d) $I_{nB}$ in the whole energy spectrum as a function of $W$  under OBCs.}
	}\label{fig6}
\end{figure}

For the nonreciprocal case in the non-Hermitian model with $\gamma=0.75$, $\Gamma=0$, and $m=3$, the result of $\bar{I}$ in Fig. \ref{fig5}(c) indicates that the localization of this non-Hermitian system is sensitive to the boundary condition. The value of $\bar{I}$ under OBCs is nonzero even in the clean limit $W=0$ due to the non-Hermitian skin effect \cite{SYao2018,HJiang2019}, where most bulk states prefer to localize in one corner of the 2D lattice [see Fig. \ref{fig6}(d) for an example]. Figure \ref{fig5}(d) shows the results of $I_n$ in the whole energy spectrum as a function of the disorder strength $W$. It is clear that $I_n$s of some eigenstates are non-zero even in the clean limit $W=0$ and are localized. For the non-Hermitian gain-and-loss case with $\gamma=0$, $\Gamma=0.25$, and $m=2.2$ in the model, the corresponding disorder-averaged IPR in Figs. \ref{fig5}(e) and \ref{fig5}(f) are similar to the Hermitian results in Figs. \ref{fig5}(a) and \ref{fig5}(b). This indicates that the non-Hermitian gain-and-loss just slightly affect the localization properties of the system under OBCs and PBCs.

{\color{black}In order to further understand the influence of the two kinds of non-Hermiticities on the localization properties, we calculate the disorder-averaged biorthogonal IPR defined under biorthogonal eigenstates~\cite{ZGong2018,PWang2019,XWLuo2019}
\begin{equation} \label{BIPR}
I_{nB}=\overline{\frac{\sum_{\textbf{x}}|\tilde{\phi}_\textbf{x}^{(s,n)}|^2}{(\sum_{\textbf{x}}|\tilde{\phi}_\textbf{x}^{(s,n)}|)^2}},
\end{equation}
where $ \tilde{\phi}_\textbf{x}^{(s,n)}={_s}\bra{nL}\textbf{\^{x}}\ket{nR}_s$. The mean biorthogonal IPR is then given by $\bar{I}_B=\frac{1}{N}\sum_{n=1}^NI_{nB}$ with $N=2L^2$. For the nonreciprocal and gain-and-loss cases, we calculate the (mean) biorthogonal IPR with the results shown in Fig. \ref{fig6}. For the nonreciprocal case with $\gamma=0.75$, $\Gamma=0$, and $m=3$ shown in Figs. \ref{fig6}(a) and \ref{fig6}(b), the values of (mean) biorthogonal IPR under OBCs are small for weak disorders, which is different from those of the right-eigenstate (mean) IPR in Figs. \ref{fig5}(c) and \ref{fig5}(d). This is due to the fact that the biorthogonal density distributions under OBCs do not suffer the skin effects \cite{PWang2019,XWLuo2019}. The right and left eigenstates under OBCs suffer the non-Hermitian skin effect, which can be understood from the similarity transformations $\ket{\tilde{n}}_s=S^{-1}\ket{nR}_s$ and $\ket{\tilde{n}}_s=S^{-1}\ket{nL}_s$ with $\ket{\tilde{n}}_s$ being the eigenstates of the corresponding Hermitian system. The corresponding biorthogonal density distribution is $ \tilde{\phi}_\textbf{x}^{(s,n)}={_s}\bra{\tilde{n}}S^{-1}\textbf{\^{x}}S\ket{\tilde{n}}_s={_s}\bra{\tilde{n}}\textbf{\^{x}}\ket{\tilde{n}}_s$, which is free of the non-Hermitian skin effect. Under PBCs, due to the absence of the non-Hermitian skin effect in this case, the biorthogonal IPR in Fig. \ref{fig6}(a) is almost the same as the IPR of right eigenstates shown in Fig. \ref{fig5}(c). For the non-Hermitian gain-and-loss case, there is no non-Hermitian skin effect of right and left eigenstates under both OBCs and PBCs. The corresponding biorthogonal IPR shown in Figs. \ref{fig6}(c) and \ref{fig6}(d) are similar to the IPR of right eigenstates displayed in Figs. \ref{fig5}(e) and \ref{fig5}(f). From these results, one can find that the IPR of right eigenstates and the biorthogonal IPR have similar localization properties in the absence of the non-Hermitian skin effect induced by the nonreciprocal hopping under OBCs, while the biorthogonal IPR does not suffer this effect.}

\begin{figure}
	\centering
	\includegraphics[width=0.48\textwidth]{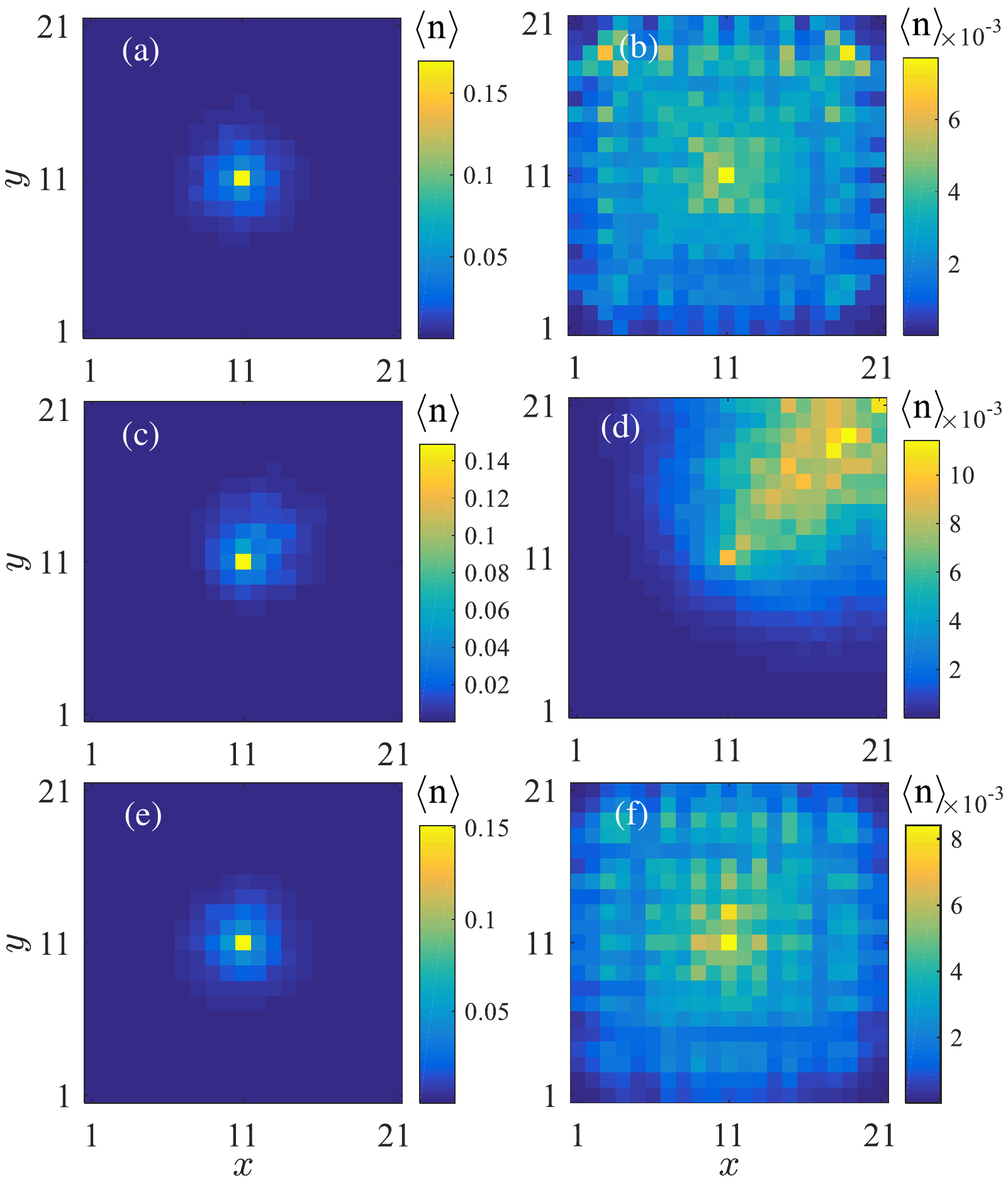}
	\caption{(Color online) The averaged density distribution of the time-evolved state $\ket{\phi(t=10)}$ with the initial state $\ket{\phi(t=0)}$ being a single particle placed at the center cell of the 2D lattice. The parameters are chosen to be the Hermitian limit with $\gamma=0$, $\Gamma=0$, and $m=2.2$ for (a) $W=2.5$ and (b) $W=0.5$; the nonreciprocal case of the non-Hermitian model with $\gamma=0.25$, $\Gamma=0$ and $m=3$ for (c) $W=3.2$ and (d) $W=1.25$, respectively; the gain-and-loss case of the non-Hermitian model with $\gamma=0$, $\Gamma=0.25$ and $m=2.2$ for (e) $W=2.5$ and (f) $W=0.5$, respectively. (a), (c), (e) and (b), (d), (f) correspond to topological nontrivial and trivial regions (see phase diagrams in Figs. \ref{fig2}, \ref{fig3} and \ref{fig4}).
	}\label{fig7}
\end{figure}

To further understand the localization properties in different topological regions, we also study the expansion of a single particle initialized at the center cell of the 2D lattice under OBCs~\cite{Lahini2010}. The initial state is prepared as $\ket{\phi(t=0)}=\frac{1}{\sqrt{2}}(c_{\textbf{x}',\uparrow}+c_{\textbf{x}',\downarrow})\ket{0}$ with the system size $L=21$ and $\textbf{x}'=((L+1)/2, (L+1)/2)$. The evolution of the initial state is ruled by the Schr\"{o}dinger equation $i\partial_t\ket{\phi(t)}=H\ket{\phi(t)}$, and the intensity of the time-dependent state $\ket{\phi(t)}$ is normalized explicitly for the non-Hermitian Hamiltonian $H$ where the evolution is no longer unitary. Figure \ref{fig7} shows the density distribution of $\ket{\phi(t=10)}$ for the Hermitian and two kinds of non-Hermitian cases, which is averaged over 100 disorder realizations. For strong disorders in the topological nontrivial regions [Figs. \ref{fig7}(a), \ref{fig7}(c), and \ref{fig7}(e)] with localized bulk states, the propagation of the particle is highly impeded by the disorder, and the density distribution is localized around the center cell (i.e., the initial position). In contrast, the distribution probability of the particle spreads into the bulk under weak disorders in the topologically trivial cases in Figs. \ref{fig7}(b), \ref{fig7}(d), and \ref{fig7}(f). Notably, the distribution probability of the particle shown in Fig. \ref{fig7} (d) reveals the tendency of the expansion into the top right-hand corner of the 2D lattice due to the non-Hermitian skin effect \cite{SYao2018,FSong2019,Kunst2018}. The interplay between disorder and nonreciprocal non-Hermiticity can also be observed in the single-particle expansion dynamics.
Note that when $\gamma$ is large enough, the density probability tends to congregate in one corner even under strong disorder and the skin effect rather than the localization being revealed by the density distribution (not shown here).

\section{\label{sec5}Conclusion and discussion}

In summary, we have investigated the interplay of topology and disorder in the 2D disordered Chern-insulator model with two types of non-Hermiticities from the nonreciprocal hopping and on-site gain-and-loss effects. We have calculated the topological phase diagrams with the open-bulk Chern number and Bott index as two topological invariants in real space. Based on the numerical results, we have revealed that the nonreciprocal hopping can enlarge the topological regions and the disorder-induced TAI can exist under both kinds of non-Hermiticities. Moreover, we have studied the localization properties of the 2D system by calculating the inverse participation ratio and the expansion evolution of a single particle.

The findings in this paper can be applied to other 2D topological models, such as the Hofstadter model and the Haldane model with additional disorder and non-Hermiticities. The three-dimensional TAI \cite{HMGuo2010} may also be extended to the non-Hermitian cases, where the interplay of the mobility edge and non-Hermitian topology may give rise to unknown physics. {\color{black} The interplay of disorder and topology could be further explored in non-Hermitian higher-order topological insulators \cite{Yuce2019,TLiu2019,Lee2019}, such as in a non-Hermitian 2D SSH model \cite{Yuce2019}.} In addition, it would be interesting to consider the cases when the disorder becomes non-Hermitian, which may lead to a TAI induced by purely non-Hermitian disorders \cite{XWLuo2019}. In future work, feasible schemes for experimental realization and detection of the proposed non-Hermitian TAI with some artificial systems will be studied.

\begin{acknowledgments}
We thank Prof. S.-L. Zhu for helpful discussions. This work was supported by the NKRDP of China (Grant No. 2016YFA0301800), the NSAF (Grant No. U1830111 and No. U1801661), the Key-Area Research and Development Program of Guangdong Province (Grant No. 2019B030330001), and the Key Program of Science and Technology of Guangzhou (Grant No. 201804020055).
\end{acknowledgments}

%

\end{document}